\documentclass[12pt,a4paper]{article}
\usepackage{latexsym}
\begin{document}

\begin{flushright}
CERN--TH/96--357\\
{\bf hep-th/9612142}\\
December $12$th, $1996$
\end{flushright}

\begin{center}


\vspace{.5cm}

{\large {\bf Extremality Versus Supersymmetry in Stringy Black Holes}}

\vspace{.9cm}

{\large
{\bf Tom\'as Ort\'{\i}n}
\footnote{E-mail address: {\tt Tomas.Ortin@cern.ch}}\\
\vspace{.4cm}
{\it C.E.R.N.~Theory Division}\\
{\it CH--1211, Gen\`eve 23, Switzerland}\\
}

\vspace{.8cm}


{\bf Abstract}

\end{center}

\begin{quotation}

\small

We study general black-hole solutions of the low-energy string
effective action in arbitrary dimensions using a general metric that
can describe them all in a unified way both in the extreme and
non-extreme cases.

We calculate the mass, temperature and entropy and study which
relations amongst the charges and the mass lead to extremality. We
find that the temperature always vanishes in the extreme limit and we
find that, for a set of $n$ charges (no further reducible by duality)
there are $2^{(n-1)}$ combinations of the charges that imply
extremality. Not all of these combinations can be central charge
eigenvalues and, thus, there are in general extreme black holes which
are not supersymmetric (or ``BPS-saturated'').

In the $N=8$ supergravity case we argue that the existence of roughly
as many supersymmetric and non-supersymmetric extreme black holes
suggests the existence of an underlying twelve-dimensional structure.

\end{quotation}

\vspace{.5cm}

\begin{flushleft}
CERN--TH/96--357\\
\end{flushleft}

\newpage

\pagestyle{plain}


\section{Introduction}

Black holes are some of the most interesting objects that string
theory seems to describe and are also the natural testing ground for
many of the new ideas in this field.  The classical black-hole
solutions of the low-energy string effective action play a central
role in these recent developments. Particularly interesting are
supersymmetric (or ``BPS-saturated'') black holes. They saturate one
or several supersymmetry bounds and describe the gravitational,
electromagnetic and other fields of states which in the quantum theory
have masses and charges protected from renormalization. The
non-renormalization theorems ensure the validity of the results
obtained in their study when they are extrapolated to the strong
coupling regime by using duality.

Since unbroken supersymmetry relates the different bosonic fields,
supersymmetric black-hole solutions are also simpler, given in terms
of a smaller number of independent functions and easier to obtain. It
seems that all supersymmetric black holes can be described in similar
terms irrespectively of the dimensionality and the number of Abelian
vector fields and scalars involved, at least in the {\it irreducible
  case} which we are going to describe now. We will say that a family
of black hole solutions is irreducible when the number of independent
electric or magnetic charges cannot be reduced by duality
transformations. The biggest irreducible family of solutions is a {\it
  generating solution} in the language of Ref.~\cite{kn:CY}.  Then,
the $n$ vector fields of an irreducible family of solutions can always
be written in this way\footnote{Here we consider the electric case
  since it is the only relevant in arbitrary number of dimensions for
  point-like objects.  In four dimensions magnetic charges can be
  described similarly in terms of the dual vector field $t$-component
  $\tilde{A}^{(i)}_{t}$.}:

\begin{equation}
A_{(i)\ t} =\alpha_{i} H_{i}^{-1}\, , 
\hspace{1cm}
\alpha_{i}=\pm 1\, ,
\end{equation}

\noindent where the $H_{i}$s are arbitrary harmonic functions 
in the relevant Euclidean transverse space. If they are to describe
black holes, we have to choose them with point-like singularities
which correspond to the different black-hole horizons (these solutions
can describe many ($N$) black holes in static equilibrium, another sign of
supersymmetry) and so, in $d$ dimensions

\begin{equation}
H_{i} = 1 +\sum_{a=1}^{N}
\frac{h_{i,a}}{|\vec{x}-\vec{x}_{a}|^{d-3}}\, .  
\end{equation}

The constants $h_{i,a}$ are all taken to be positive to avoid naked
singularities (why this is so will become clear when we say how these
functions enter the metric). They are otherwise arbitrary. Then, since
 black holes can carry positive or negative charge for each
$U(1)$, the constants $\alpha_{i}$ must be allowed to be positive or
negative. The $i$th $U(1)$ charge of the $a$th black hole is
proportional to the product $\alpha_{i}h_{i,a}=\pm h_{i,a}=q_{i,a}$
and so all black holes have charges with the same sign for each $U(1)$
and $\alpha_{i}={\rm sign}\ (q_{i,a})$.  The harmonic functions are
normalized to be $1$ at infinity so we do not need to introduce any
other constants in order to have an asymptotically flat metric.

The Einstein-frame  metric is always of this form:

\begin{equation}
\label{eq:extrememetric}
ds^{2}=\left(\prod_{i=1}^{i=n} H_{i}^{-2r_{i}}\right) dt^{2}
-\left(\prod_{i=1}^{i=n} H_{i}^{-2r_{i}}\right)^{-\frac{1}{d-3}}
d\vec{x}^{\ 2}\, ,
\end{equation}

\noindent where $\vec{x}=(x^{1},\ldots,x^{d-1})$.

It is a experimental fact that the coefficients $r_{i}$ always satisfy

\begin{equation}
\label{eq:property}
\sum_{i=1}^{i=n}r_{i}=1\, ,  
\end{equation}

\noindent and when they do not, we can consider that there is an
additional vector field with respect to which the black holes are
uncharged, so the corresponding $h_{i,a}=0$ and the harmonic function
is simply unity and cannot be ``seen'' in Eq.~(\ref{eq:extrememetric})
even though it is actually there\footnote{This is what happens, for
  instance, in the $U(1)$ dilaton black holes with $a=1$ of
  Refs.~\cite{kn:GGM,kn:GHS}. The extreme black holes of this family,
  which are also supersymmetric \cite{kn:KLOPP} can be described in
  terms of a single harmonic function with coefficient $r=1/2$.
  However they can be considered as a particular case of extreme
  dilaton $U(1)\times U(1)$ black holes which can be described in
  terms of two harmonic functions with $r_{1}=r_{2}=1/2$ so $\sum_{i}
  r_{i}=1$. The $U(1)$ black holes are obviously those for which one
  charge vanishes and the corresponding harmonic function is just
  $1$.}. So we can assume that Eq.~(\ref{eq:property}) always holds
true in a certain coordinate system when all possible charges are
switched on.

In most cases one also has that all the $r_{i}$s are equal. If they
are not, one can assume that there are more harmonic functions but
they happen to be equal for the particular solution one is
considering. For instance, if there are two harmonic functions with
$r_{1}=1/3,\,\, r_{2}=2/3$ one can consider that in the most general
case there are three harmonic functions with all $r_{i}$s equal to
$1/3$ but that $H_{2}=H_{3}$. This assumption is not crucial in what
follows and we will not use it, although it should be kept in mind
that in most cases it is true.

If we take harmonic functions with a single pole in the origin, the
ADM mass $m$ of the single black hole they describe, which we define by

\begin{equation}
g_{tt} \sim 1 -\frac{2m}{\rho^{d-3}}\, ,  
\end{equation}

\noindent is given by

\begin{equation}
m= \sum_{i=1}^{i=n}r_{i}|q_{i}|\, .
\end{equation}

Thus, the mass of all the supersymmetric black holes of the above form
is related to their charges. A relation of this kind is expected
whenever a solution saturates a supersymmetry (or Bogomol'nyi) bound.
However, as we are going to see, a relation of this kind does not
guarantee the supersymmetry of the corresponding solution.

In all known cases, {\it supersymmetric} (static) black holes are also
{\it extreme} black holes. This means that they can be obtained by
taking the appropriate limit of a family of charged black hole
solutions.  These, in general have two horizons: an event horizon and
a Cauchy horizon. When the mass diminishes these horizons get closer,
eventually coinciding in a single event horizon. This is the extreme
limit.

All extreme black holes of the low-energy effective string theory can
be written in the above form but not all of them are supersymmetric.
A extreme black hole which is not supersymmetric has been known for
some time\footnote{Obviously there are many extreme dilaton black hole
  which are extreme but not supersymmetric because they cannot be
  considered solutions of any supergravity theory, to start with
  \cite{kn:KhO1}. } \cite{kn:G} (see also \cite{kn:KhO2}). One of our
goals in this letter will be to show that this is actually a very
general phenomenon.

It turns out that the general (non-extreme) black holes of the
low-energy string theory can be found by modifying the above solutions
with the introduction of an extra factor $W$ in the metric which we
henceforth refer to as the {\it Schwarzschild factor}\footnote{This
  procedure was first used in Ref.~\cite{kn:CT} and further exploited
  in Refs.~\cite{kn:C,kn:O}}:

\begin{equation}
\label{eq:nonextrememetric}
ds^{2}=\left(\prod_{i=1}^{i=n} H_{i}^{-2r_{i}}\right) Wdt^{2}
-\left(\prod_{i=1}^{i=n} H_{i}^{-2r_{i}}\right)^{-\frac{1}{d-3}}
\left[W^{-1}d\rho^{2} +\rho^{2}d\Omega^{2}_{(d-2)} \right]\, ,
\end{equation}

\noindent where the Schwarzschild factor is given by

\begin{equation}
W = 1 - \frac{2r_{0}}{\rho^{d-3}}\, ,  
\hspace{1cm}
r_{0}>0\, ,
\end{equation}

\noindent and the old harmonic functions are now forced to be of the
form

\begin{equation}
H_{n}=1 + \frac{h_{n}}{\rho^{d-3}}\, .  
\end{equation}

There are two more modifications: the constants $\alpha_{i}$ are no
longer simply $\pm 1$ but are related to the $h_{i}$s and to $r_{0}$.
The relation is always such that taking into account that the charges
are $q_{i}\sim \alpha_{i}h_{i}$ one can write

\begin{equation}
h_{i} =-r_{0} +\sqrt{r_{0}^{2} +q_{i}^{2}}\, .  
\end{equation}

Observe that when the Schwarzschild factor $W=1$ we recover the
extremal single black hole solutions. For this reason $r_{0}$ is
called the {\it extremality parameter}. In fact, all is used in
checking that the above are solutions is the equation

\begin{equation}
\label{eq:hi}
\partial_{\rho}\left(\rho^{d-3}\partial_{\rho}H_{i} \right) =0\, ,  
\end{equation}

\noindent so when $W=1$ one can go to Cartesian coordinates in which 
the above equation tells us simply that the $H_{i}$s are arbitrary
harmonic functions. Therefore, the above solutions include the extreme
black-hole solutions in the $r_{0}\rightarrow 0$ limit.

Our goal is to study general black-hole solutions of this kind trying
to derive as many general consequences about their physical properties
as possible.


\section{Some Examples}

It is worth seeing how all this is realized in a few simple examples.
The four-dimensional Reissner-Nordstr\"{o}m black-hole solution is
usually written in this form:

\begin{eqnarray}
ds^{2} & = & \frac{(r-r_{+})(r-r_{-})}{r^{2}}dt^{2}
-\frac{r^{2}}{(r-r_{+})(r-r_{-})} dr^{2} -r^{2}d\Omega^{2}\, , \nonumber \\
& & \\
A_{\mu} & = & -\delta_{\mu t}\frac{Q}{r}\, , \nonumber
\end{eqnarray}

\noindent where the integration constants $r_{\pm}$ are given in terms
of the electric charge $Q$ and the ADM mass M by

\begin{equation}
r_{\pm}= M \pm r_{0}\, ,
\hspace{1cm}
r_{0}^{2} =M^{2}-Q^{2}\, .
\end{equation}

If we shift the radial coordinate by $r=\rho +r_{-}$ and we make a
gauge transformation of the gauge potential that results into the
addition of the constant $Q/r_{-}$, the it takes the following form:

\begin{eqnarray}
ds^{2} & = & \left( 1 +\frac{r_{-}}{\rho} \right)^{-2}
\left( 1 - \frac{2r_{0}}{\rho} \right) dt^{2}
\nonumber \\
& & \nonumber \\
& &
-\left( 1 +\frac{r_{-}}{\rho} \right)^{2} \left[
\left( 1 - \frac{2r_{0}}{\rho} \right)^{-1} d\rho^{2}
+\rho^{2}d\Omega^{2}\right]\, , \nonumber \\
& & 
\label{eq:RN2} \\
A^{\prime}_{\mu} & = & \delta_{\mu t}\frac{Q}{r_{-}}
\left( 1 + \frac{r_{-}}{\rho} \right)^{-1}\, . \nonumber 
\end{eqnarray}

Obviously here we have 

\begin{equation}
H =  1 +\frac{r_{-}}{\rho}\, ,
\hspace{1cm}
\alpha= \frac{Q}{r_{-}}\, .
\end{equation}

The higher-dimensional Reissner-Nordstr\"om solutions can be also
written in this form. Other solutions are the dilaton black-hole
solutions of the $a$-model (which include the Reissner-Nordstr\"om
solutions) in $d$ dimensions with action

\begin{equation}
S = \frac{1}{16\pi G_{N}^{(d)}} \int d^{d}x\sqrt{|g|} \left[R 
+2(\partial\varphi)^{2}  -e^{-2a\varphi}F^{2}\right]\, ,
\label{eq:amodelaction}
\end{equation}

\noindent which can be written as follows:

\begin{equation}
\left\{
\begin{array}{rcl}
ds^{2} & = & \left(e^{-2a(\varphi-\varphi_{0})} H^{-2}\right) 
W dt^{2}  \\
& & \\
& & 
-\left(e^{-2a(\varphi-\varphi_{0})} H^{-2}\right)^{\frac{-1}{d-3}}
\left[W^{-1}d\rho^{2} +\rho^{2}d\Omega_{d-2}^{2}  \right]\, , \\
& & \\
A_{\mu} & = & \alpha \delta_{\mu t} H^{-1}\, , \\
& & \\
e^{-2a\varphi} & = & e^{-2a\varphi_{0}} H^{2k}\, , \\
& & \\
H & = & 1+\frac{h}{\rho^{d-3}}\, , 
\hspace{1cm}
W=1-\frac{2r_{0}}{\rho^{d-3}}\, , \\
& & \\
2r_{0} & = & -h\left[1 -\frac{a^{2}}{x}\alpha^{2} \right]\, , \\
& & \\
k & = &  \frac{a^{2}}{2} c/(1+\frac{a^{2}}{2}c)\, ,
\hspace{1cm}
c=(d-2)/(d-3)\, . \\
\end{array}
\right.
\end{equation}

Although our assumption for the constants $r_{i}$ does not always seem
to be true in this case, one should not forget that most of the solutions
of this model are not solutions of any supergravity theory.

A case in which there are two harmonic functions is given by the
solutions to the $d$-dimensional $a_{1}-a_{2}$-model with action

\begin{equation}
S = \frac{1}{16\pi G_{N}^{(d)}} \int d^{d}x\sqrt{|g|} \left[R 
+2(\partial\varphi)^{2}  -e^{-2a_{1}\varphi}F_{(1)}^{2}
-e^{-2a_{2}\varphi}F_{(2)}^{2}\right]\, .  
\label{eq:a1a2BHmodelaction}
\end{equation}

When $a_{1}$ and $a_{2}$ are related by

\begin{equation}
\label{eq:constraint}
a_{1}a_{2} = -2  \left(\frac{d-3}{d-2} \right)\, .
\end{equation}

\noindent one finds the following solutions \cite{kn:GGM}:

\begin{equation}
\left\{
\begin{array}{rcl}
ds^{2} & = & \left(e^{-2a_{1}(\varphi-\varphi_{0})}H_{1}^{-2}\right)
 Wdt^{2} \\
& & \\
& &
-\left(e^{-2a_{1}(\varphi-\varphi_{0})}H_{1}^{-2}\right)^{-\frac{1}{d-3}}
\left[W^{-1} d\rho^{2} +\rho^{2}d\Omega_{(d-2)}^{2}\right]\, , \\
& & \\
e^{-2\varphi} & = & 
e^{-2\varphi_{0}}\left(H_{1}/H_{2} \right)^{\frac{2}{(a_{1}-a_{2})}}\, . \\
& & \\
A_{(i) \mu} & = & \delta_{\mu t} \alpha_{i} e^{a_{i}\varphi_{0}} 
H_{i}^{-1}\, ,\\
& & \\
H_{i} & = & 1 +h_{i}/\rho^{d-3}\, ,
\hspace{.5cm}
W  =  1 -2r_{0}/\rho^{d-3}\, , \\
& & \\
2r_{0} & = & -h_{1} \left[ 1 -a_{1}(a_{1}-a_{2})\alpha_{1}^{2}\right]\, ,
\hspace{.5cm}
2r_{0}  =  -h_{2} \left[ 1 +a_{2}(a_{1}-a_{2})\alpha_{2}^{2}\right]\, .
\end{array}
\right.
\end{equation}

There are only two cases (known to us) in which this model really
arises as a truncation of a string theory action: in $d=5$ dimensions,
where $F_{(2)}$ would be the dual of the NS-NS 3-form (the axion field
strength) and we would have $a_{1}=2/\sqrt{6},a_{2}=-4/\sqrt{6}$, and
in $d=4$ dimensions where $F_{(2)}$ would be a RR vector field and we
would have $a_{1}=+1,a_{2}=-1$.  The problem in higher dimensions is
that $F_{(2)}$ would have to be the dual of a 4-form, 5-form etc. and
all these fields are RR-type fields and, in the string frame, where
the dilaton is an overall factor for the NS-NS fields, the RR fields
do not couple directly to it and, in the Einstein frame, the
constraint (\ref{eq:constraint}) is not satisfied. Of course, other
possibilities arise if $\varphi$ is not considered to be the dilaton
but some other moduli fields or combinations of them. In any case, in
the two stringy cases of interest the solution nicely fits into the
general metric written in the previous section.

More black-hole solutions with this structure can be found elsewhere
\cite{kn:CT,kn:C}.


\section{Physical Properties}


\subsection{Extremality and Supersymmetry}

The independent parameters of the charged black-hole solutions are the
ADM mass $m$ and the charges $q_{i}$ which are, up to factors that we
absorb in their definition, $h_{i}\alpha_{i}$. In these factors are
included all moduli.

The first thing we would like to do is to express all the parameters
that appear in the solution in terms of the physical parameters. The
$h_{i}$s are expressed in terms of $r_{0}$ and the charges $q_{i}$
through the equations (\ref{eq:hi}). Then we only need to express
$r_{0}$ in terms of the physical parameters. If we calculate the mass
of the black-hole metric (\ref{eq:nonextrememetric}) we get

\begin{equation}
\label{eq:r0def}
m = \sum_{i=1}^{i=n}r_{i}\sqrt{r_{0}^{2} +q_{i}^{2}}\, .
\end{equation}

This equation implicitly gives $r_{0}$ as a function of the mass and
charges. It is impossible to solve for $r_{0}$ in the general case but
we can extract very interesting information from the above equation. 

First, let us see how it can be solved in the simplest cases. We set
$r_{i}=1/n$. For $n=1$ it is trivial to get 

\begin{equation}
r_{0}^{2} = m^{2} -q^{2}\, .
\end{equation}

$r_{0}$ vanishes when $m=|q|$ and, in principle, we could identify the
central charge of $N=2$ supergravity (or any of the central charge
matrix skew eigenvalues in higher  $N=4,8$ supergravity \cite{kn:FSZ})
with $q$  so $|z|=|q|$. This, indeed seems to be always the case.

For $n=2$ the above equation can be transformed in a quadratic
algebraic equation for $r_{0}^{2}$ whose solution is

\begin{equation}
r_{0}^{2} = \frac{1}{m^{2}} 
\left[m^{2} -\left(\frac{q_{1}+q_{2}}{2}\right)^{2}\right]
\left[m^{2} -\left(\frac{q_{1}-q_{2}}{2}\right)^{2}\right]\, .
\end{equation}

We see that $r_{0}$ vanishes when $m^{2}$ equals one of the combinations

\begin{equation} 
\begin{array}{rcl}
|z_{1}| & = & \left|\frac{q_{1}+q_{2}}{2}\right|\, , \\
& & \\
|z_{2}| & = & \left|\frac{q_{1}-q_{2}}{2}\right|\, , \\
\end{array}
\end{equation}

\noindent which can be identified with the two central charge skew
eigenvalues  of $N=4$ supergravity or two of the four skew eigenvalue of
the central charge matrix of $N=8$ supergravity. 

For $n=3$ one gets a quartic algebraic equation for $r_{0}^{2}$. The
general solution is complicated but we can learn the same from a
simplified case: $q_{2}=q_{3}$. (This is the case  that applies to
black-hole solutions of pure $d=5,N=4$ supergravity, which is a
consistent truncation of the low-energy string effective action.) The
solution is

\begin{equation}
\label{eq:r0n3}
r_{0}^{2} = m^{2} + \frac{q_{1}^{2}-q_{2}^{2}}{3} +4m^{2} -q_{2}^{2}
-4m \left(m^{2} + \frac{q_{1}^{2}-q_{2}^{2}}{3} \right)^{\frac{1}{2}}\, .
\end{equation}

We immediately notice that the expression for $r_{0}$ does not
factorize into ``Bogomol'nyi factors'' $(m^{2}-|z_{i}|^{2})$ as it
happened in the $n<3$ cases. It is easy to see that $r_{0}$ vanishes
when $m^{2}$ equals either of the two combinations

\begin{equation} 
\begin{array}{rcl}
|z_{1}| & = & \left|\frac{q_{1}+2q_{2}}{3}\right|\, , \\
& & \\
|z_{2}| & = & \left|\frac{q_{1}-2q_{2}}{3}\right|\, , \\
\end{array}
\end{equation}

\noindent which can be naturally identified with the skew eigenvalues of
the central charge matrix of $N=4$ supergravity. These two combinations
do not appear in Eq.~(\ref{eq:r0n3}), which is understandable because
$r_{0}$ is not a polynomial on $m^{2}$.

For $n\geq4$ Eq.~(\ref{eq:r0def}) cannot be rewritten as a polynomial
on $r^{2}_{0}$. It is, therefore, impossible to write $r_{0}$ as a
product of Bogomol'nyi factors in all the cases with $n\geq 3$. This
applies, in particular, for the black-hole solutions of the heterotic
string effective action in four and five dimensions ($n=4,3$
respectively) and those of the type~II theory in the same dimensions.

In spite of the difficulty of finding a expression for the extremality
parameter in terms of the mass and charges, it is easy to find from
the defining equation (\ref{eq:r0def}) the zeroes of $r_{0}$. When
$r_{0}=0$, the mass and charges are related by

\begin{equation}
\label{eq:r0zeroes}
m = \sum_{i=1}^{i=n}r_{i}|q_{i}|\, .
\end{equation}

We would like to rewrite this relation in a Bogomol'nyi-like fashion,
$m=|z|$ which is a necessary condition for the corresponding extreme
black holes to be supersymmetric\footnote{It is not a sufficient
  condition because we would have to prove that the $|z|$s so obtained
  are the actual central charge skew eigenvalues, but it is very
  reasonable to expect it.}. Is is easy to see that the above equation
can be rewritten in $2^{(n-1)}$ different Bogomol'nyi-like ways:

\begin{equation}
m  =  |\sum_{q_{i}>0}r_{i} q_{i} -\sum_{q_{i}<0}r_{i} q_{i}|\, , 
\end{equation}

\noindent which correspond to the $2^{(n-1)}$ possible relative signs 
among the charges. In general, this number of Bogomol'nyi-like
identities is clearly larger than the number of skew eigenvalues of
the central charge matrix. To be specific, in pure $N=2,4$
supergravity in $d=4,5$ dimensions this problem does not occur.
However, in $N=8$ supergravity we would have found $8$ of these
Bogomol'nyi-like identities while there are only $4$ supersymmetry
bounds. Then, in this theory, there are extreme black holes which
satisfy the identity (\ref{eq:r0zeroes}) by satisfying one of the four
Bogomol'nyi-like identities which are not supersymmetry bounds and
therefore are not supersymmetric.

This can be better seen in an example. The black-hole generating
solutions of the heterotic compactified in a six-torus can be found as
solutions of the truncated action

\begin{eqnarray}
S & = & \int dx^{4}\ \sqrt{|g|} \left\{ R
+2\left[ (\partial \phi)^{2} +(\partial \sigma)^{2}
+(\partial \rho)^{2}\right] \right.
\nonumber \\
& & \nonumber \\
& &
-{\textstyle\frac{1}{4}} e^{-2\phi}
\left[
e^{-2(\sigma+\rho)} (F_{1})^{2}
+e^{-2(\sigma-\rho)} (F_{2})^{2}
\right.
\nonumber \\
& & \nonumber \\
& &
\left.
\left.
+e^{2(\sigma+\rho)} (F_{3})^{2}
+e^{2(\sigma-\rho)} (F_{4})^{2} \right] \right\}\, .
\end{eqnarray}

These solutions were found in Ref.~\cite{kn:CY2}, further discussed in
Refs.~\cite{kn:CT2} and later rediscovered in Ref.~\cite{kn:R}.

The extreme solution is given in terms of four independent harmonic
functions $H_{i}$, $i=1,\ldots,4$

\begin{equation}
\begin{array}{rcl}
ds^{2} & = & \left( \prod_{i=1}^{i=4}H_{i}^{-\frac{1}{2}}\right) dt^{2} 
- \left( \prod_{i=1}^{i=4}H_{i}^{-\frac{1}{2}}\right)^{-1} d\vec{x}^{\ 2}
\, ,\\
& & \\
e^{-4\phi} & = & \frac{H_{1}H_{3}}{H_{2}H_{4}}\, ,
\hspace{1cm}
e^{-4\sigma} =\frac{H_{1}H_{4}}{H_{2}H_{3}}\, ,
\hspace{1cm}
e^{-4\rho} =\frac{H_{1}H_{2}}{H_{3}H_{4}}\, ,
 \\
& &  \\
A_{(i)\ t} & = & \alpha_{i}H_{i}^{-1}\, ,\,\, i=1,3\, , 
\hspace{.5cm}
\tilde{A}_{(i)\ t}  =  \alpha_{i}H_{i}^{-1}\, ,\,\, i=2,4\,  
\end{array}
\end{equation}

\noindent where $\alpha_{i}={\rm sign }\ (q_{i})= \pm 1$ and

\begin{equation}
\tilde{F}_{2} =e^{-2(\phi+\sigma-\rho)}
{}^{\star}F_{2}\, ,
\hspace{1cm}
\tilde{F}_{4} =e^{-2(\phi-\sigma+\rho)}
{}^{\star}F_{4}\, ,
\end{equation}

\noindent and ${}^{\star}F$ is the Hodge dual of $F$. 
The harmonic functions, for a single black hole are

\begin{equation}
H_{i} = 1 +\frac{|q_{i}|}{|\vec{x}|}\, .  
\end{equation}

(The charges $q_{2}$ and $q_{4}$ are here magnetic charges.)  These
solutions fit our general metric for extreme string black holes with
$n=4$ and $r_{i}=1/4$. The ADM mass is, thus, given by

\begin{equation}
m = {\textstyle\frac{1}{4}} \sum_{i=1}^{i=4}|q_{i}|\, .  
\end{equation}

The four central charge skew eigenvalues $z_{i}$ of $N=8$ supergravity
(where we can embed this solution too) are given in terms of the
charges $q_{i}$ by \cite{kn:KK}

\begin{equation}
\begin{array}{rcl}
|z_{1}| & = &  \frac{1}{4}|q_{1}+q_{2}+q_{3}+q_{4}|\, , \\
& & \\
|z_{2}| & = &  \frac{1}{4}|q_{1}-q_{2}+q_{3}-q_{4}|\, , \\
& & \\
|z_{3}| & = &  \frac{1}{4}|q_{1}+q_{2}-q_{3}-q_{4}|\, , \\
& & \\
|z_{4}| & = &  \frac{1}{4}|q_{1}-q_{2}-q_{3}+q_{4}|\, . \\
& & \\
\end{array}
\end{equation}

Now, it is easy to see that there are extreme black holes which
satisfy the mass formula but do not saturate any Bogomol'nyi bound
(essentially half of the total).  Taking unit charges, so
$m=|q_{i}|=1$, we have the following eight possibilities:

\begin{equation}
\vec{q}=\pm(1,1,1,1)\, ,
\hspace{1cm}
m=|z_{1}|=1\, ,   
\end{equation}

\noindent and the black hole has one unbroken supersymmetry. 
The next case is

\begin{equation}
\vec{q}=\pm(1,1,1,-1)\, ,
\hspace{1cm}
m={\textstyle\frac{1}{4}}|q_{1}+q_{2}+q_{3}-q_{4}|=1\, ,   
\end{equation}

\noindent while

\begin{equation}
|z_{1}| = |z_{2}| = |z_{3}| = |z_{4}| = {\textstyle\frac{1}{2}}<m\, ,
\end{equation}

\noindent and the black hole has no unbroken supersymmetries in spite
of being extreme. Observe that this black hole satisfies the four
supersymmetry bounds without saturating any of them. The same happens
in the other three cases in which one charge has sign different to the
other three. The next case is

\begin{equation}
\vec{q}=\pm(1,1,-1,-1)\, ,
\hspace{1cm}
m=|z_{3}|\, ,   
\end{equation}

\noindent and the black hole has one unbroken supersymmetry, etc.


\subsection{Temperature}

It is straightforward to find that the temperature is in general given
by

\begin{equation}
T = \frac{(d-3)}{2^{\frac{1}{d-3}} 4\pi} 
\frac{r_{0}^{2-\frac{1}{d-3}}}{\prod_{i=1}^{i=n} 
\left[ r_{0} + \sqrt{r_{0}^{2} +q_{i}^{2}} \right]^{2r_{i}}}  \, ,
\end{equation}

\noindent which shows that it always vanishes in the extreme limit.
However, some care has to be taken when some of the charges vanish
because, then, there is an ambiguity in the calculation of the
temperature: there are additional $r_{0}$ factors in the denominator
which may cancel those of the numerator and then the temperature may
not vanish. The resolution of this paradox is the same as in the
four-dimensional $U(1)\times U(1)$ dilaton black holes
\cite{kn:GGM,kn:KLOPP}.


\subsection{Entropy}

The general formula for the entropy is

\begin{equation}
S = \frac{\omega_{d-2}}{4} \left\{2 \prod_{i=1}^{i=n} 
\left[ r_{0} + \sqrt{r_{0}^{2} +q_{i}^{2}} \right]^{r_{i}} 
\right\}^{\frac{d-2}{d-3}}  \, ,
\end{equation}

\noindent where $\omega_{d-2}$ is the volume of the $(d-2)$-sphere.
In the extreme limit $S$ is always proportional to a product of charges

\begin{equation}
S = \frac{\omega_{d-2}}{4} 2^{\frac{d-2}{d-3}} \prod_{i=1}^{i=n} 
|q_{i}|^{r_{i}\frac{d-2}{d-3}}\, .
\end{equation}

This formula generalizes the formulae obtained in Ref.~\cite{kn:KR}.

Only in four dimensions there is a simple relation between the
extremality parameter $r_{0}$ and the product $ST$:

\begin{equation}
ST = \frac{(d-3)\omega_{d-2}}{16\pi} 2 r_{0}^{\frac{2d-7}{d-3}}  
\left\{2 \prod_{i=1}^{i=n} 
\left[ r_{0} + \sqrt{r_{0}^{2} +q_{i}^{2}} \right]^{r_{i}} 
\right\}^{-\left(\frac{d-4}{d-3}\right)}\, .
\end{equation}


\section{Conclusion}

We have studied general black hole solutions of the low-energy string
effective action and computed their mass, temperature and entropy. we
have also studied the extremality conditions and have found that in
$N=2,4$ supergravity plus matter as well as in $N=8$ supergravity
there are many extreme black holes which do not saturate any
Bogomol'nyi bound. 

In fact, in $N=8$ supergravity roughly a half of the extreme black
holes do not saturate any of the four Bogomol'nyi bounds of this
theory and are not supersymmetric. Instead they saturate one of a set
of four different Bogomol'nyi-like bounds.

An intriguing possibility, previously suggested in Ref.~\cite{kn:KhO2}
is that $N=8$ can be considered as a consistent truncation of a $N=16$
supergravity theory which would not be consistent before the
truncation is done. A concrete scenario was proposed there which may
have a twelve-dimensional origin, in a sense similar to the one
proposed by Kutasov and Martinec's $N=(2,1)$ string scenario
\cite{kn:KM1,kn:KMO,kn:M,kn:KM2} and Bars' ``S~theory'' scenario
\cite{kn:B1,kn:B2}\footnote{By now it seems clear that F~theory
  \cite{kn:V} is not a fully twelve-dimensional theory. The existence
  of a twelve-dimensional theory has also been proposed in
  Ref.~\cite{kn:H}.}.

Such an $N=16$ theory would have $8$ central charge skew eigenvalues
and the $N=8$ would ``remember'' that four combinations of the charges
were central charge eigenvalues even though after the truncation they
are not any more. This proposal could eventually be checked in the
framework of the $N=(2,1)$ string or S~theory scenarios.


\section*{Acknowledgements}

The author is most grateful to J.L.F.~Barb\'on and R.~Kallosh for most
useful conversations and to M.M.~Fern\'andez by her support.


\end{document}